# ENHANCED GREY BOX FUZZING FOR INTEL MEDIA DRIVER

Linlin Zhang[1] and Ning Luo[2]

[1]Visual Computing Group, Intel Asia-Pacific Research & Development Ltd, Shanghai, China
livia.zhang@intel.com
[2]Visual Computing Group, Intel Asia-Pacific Research & Development Ltd, Shanghai, China
ning.luo@intel.com

### ABSTRACT

*Grey box fuzzing is one of the most successful methods for automatic vulnerability detection. However, conventional Grey box Fuzzers like AFL can open perform fuzzing against the whole input and spend more time on smaller seeds with lower execution time, which significantly impact fuzzing efficiency for complicated input types. In this work, we introduce one intelligent grey box fuzzing for Intel Media driver, MediaFuzzer, which can perform effective fuzzing based on selective fields of complicated input. Also, with one novel calling depth-based power schedule biased toward seed corpus which can lead to deeper calling chain, it dramatically improves the vulnerability exposures (~6.6 times more issues exposed) and fuzzing efficiency (~2.7 times more efficient) against the baseline AFL for Intel media driver with almost negligible overhead.*

## 1. INTRODUCTION

Grey box fuzzing is a popular and effective approach for vulnerability discovery. As opposed to black box approaches which suffer from a lack of knowledge about the application, and white box approaches which can incur high overheads due to program analysis, grey box leverage a lightweight code instrumentation approach to achieve the balance between efficiency and overheads. American Fuzzy Lop (AFL) and its variants are the most popular implementations of Grey box fuzzers.

However, several limitations may greatly impact the fuzzing efficiency for large scale software with complicated input, like Intel Media Driver. Firstly, without input structure awareness, Grey box fuzzers usually perform bit level mutation on the whole input indiscriminately which is ineffective on exploring the vast yet sparse domain of expected inputs. Secondly, they tend to spend more time on seeds with lower execution time – in real case, 99% of fuzzed inputs will be directly rejected by the input validity check and cannot enter & verify the core logic of the software.

In this work, we introduce one novel intelligent grey box fuzzing for Intel Media driver, MediaFuzzer. Which can do effective fuzzing based on selective fields of the input. As the core of MediaFuzzer, we also create one novel calling depth-based power schedule biased toward seed corpus leading to deeper calling chain and more likely to pass the parameter validity check.

Per Our evaluation, against the above two innovations, MediaFuzzer can dramatically improve the vulnerability exposures (~6.6 times more issues exposed) as well as the fuzzing efficiency (~2.7 times more efficient) comparing with its baseline AFL for Intel media driver with almost negligible extra overhead.

## 2. Media Fuzzer Overview

Media Fuzzer working flow can be shown by [Figure 1] below.

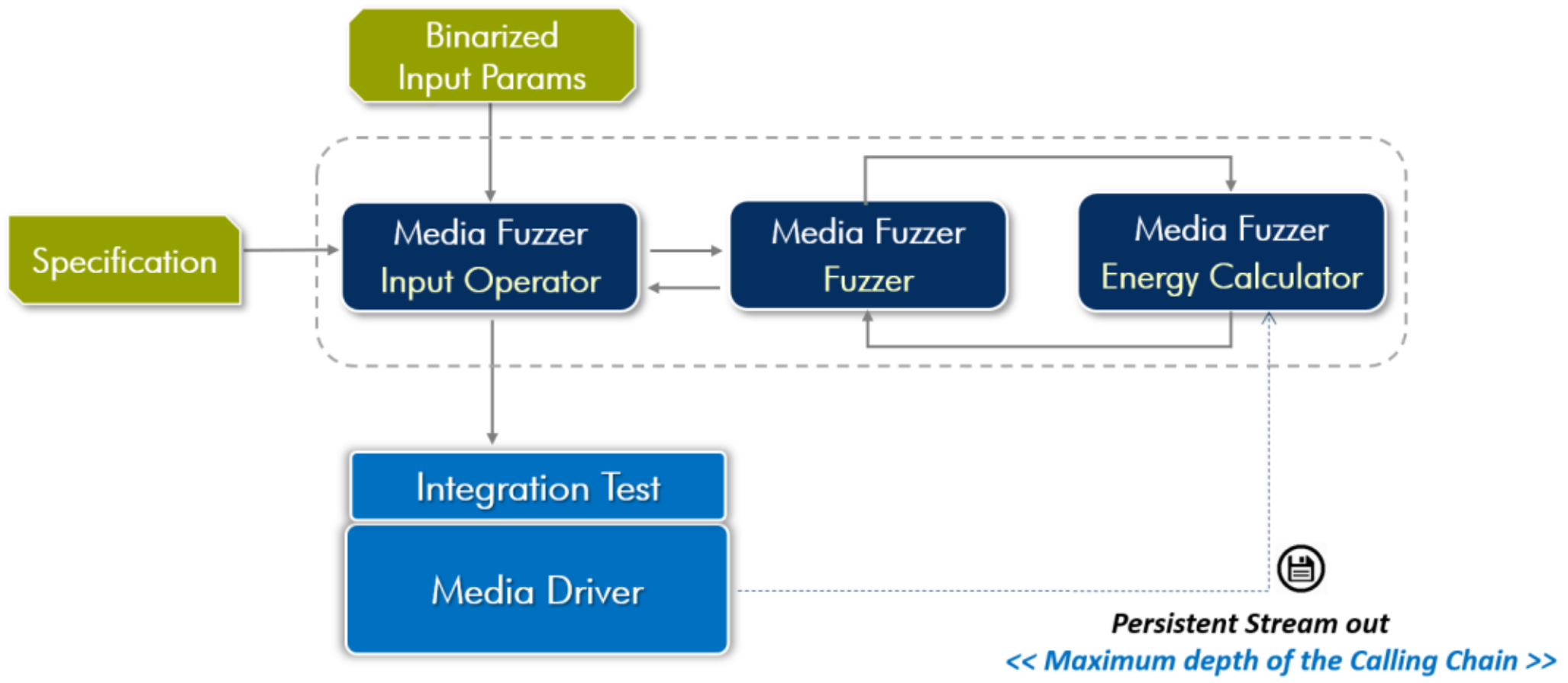

Figure 1: Media Fuzzer Working Flow

Media Fuzzer is hooked onto the input of Intel Media Integration test. The input parameters will be binarized and feed into Media Fuzzer before entering the real test case.

For each input param type, there is one accompanying specification file articulating its binary layout as well as those interesting fields to the following fuzzing operation.

Fuzzing operation will be performed against Partial of the input spliced from the aforementioned interesting bits. Input operator will parse the binarized input, extract the interesting fields and generate initial input (seed corpus), based on settings in specification file. The seed corpus generated will be used as the input to the following fuzzing pipeline.

Starting with initial input (seed corpus), Fuzzer module of MediaFuzzer, which is implemented on top of AFL core, will perform the mutation against pre-defined set of generic mutation operators (e.g., bitflips).

Energy Calculator determines how much time is spent on fuzzing operations for one seed.

In Media Fuzzer, we innovated one validity-based power schedule which tends to assign more energy to (spend more time on) inputs leading to deeper calling chain.

Based on the decision from Energy Calculator, the mutated inputs deemed sufficiently new will be mutated further to explore more inputs.

The Fuzzed input will then be pass back to Input Operator and restored to its original form and then feed into Media driver Integration test for test case execution.

Series of trace messages are added inside Intel Media Driver to record to the maximum calling chain depth (inside driver) in current execution. After each execution, one special designed postscript will help to parse the output, extract the aforementioned maximum calling chain depth data, and get it feedback to Energy Calculator for energy data update of the active seed.

Next let us put more focus on the 2 key innovations in MediaFuzzer:

- **Effective fuzzing based on Partial Input**

In contrast to conventional Grey box fuzzers, MediaFuzzer is input-structure aware which can perform bit level mutation on selective fields for effective fuzzing based on partial of the input.

As can be shown by [Figure 2] below,

Before fuzzing, all input params will be binarized and passed into Input operator module firstly. Input operator will extract those interesting fields based on the specification settings and generate the initial seed corpus for the following fuzzing operation.

After Fuzzing operation done, the fuzzed seed will be passed back to Input operator again and restored into their original form. The restored input with fuzzed bits will then be used as the input to media integration test.

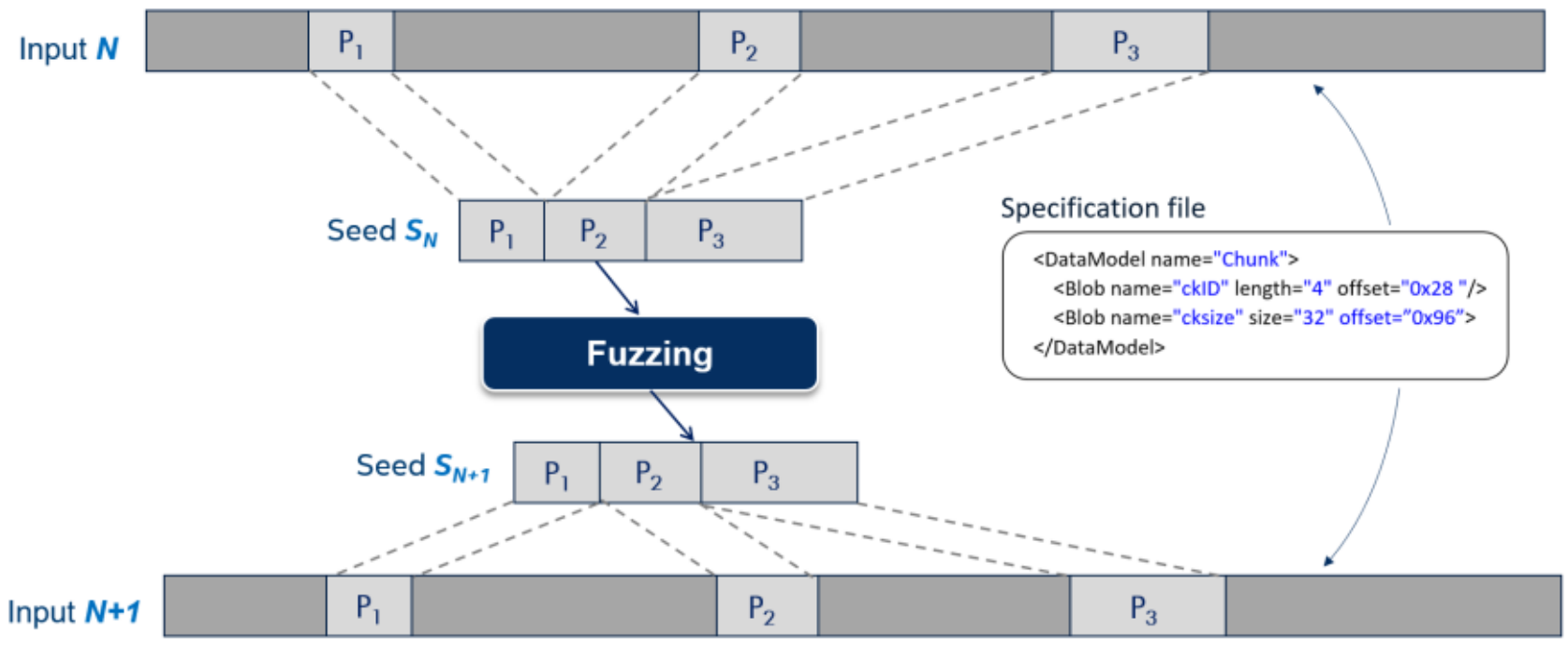

Figure 2. Enhanced fuzzing based on Partial of input

- **Power Schedule based on Calling Chain Depth**

Power schedule in fuzzer is to determine how much time should be spent on fuzzing operation for each seed. Several conventional power schedules are used before. E.g, the power schedule of AFL assigns more energy to smaller seeds with a lower execution time that have been discovered later.

For effective fuzzing on Media Driver, we introduce one novel calling chain depth-based power schedule in which each seed is assigned an energy based on the maximum depth of the calling chain in each execution (triggered by this seed). It tends to spend more time on seeds which can step deeply into the driver under test.

Some more details about the forementioned calling chain depth-based power schedule are as below:

The degree of validity v(s) of a seed s is determined by the maximum calling depth of the last execution. If last time the seed lead to the historical maximum calling depth, its degree of validity v(s) = 100%. If the previous calling depth is half of the historical maximum, its validity v(s) = 50%.

Given the seed s, the depth-based power schedule power schedule Pv(s) assigns energy as below

$$p_v(s) = \begin{cases} 2p(s) & \text{if } v(s) \geq 50\% \text{ and } p(s) \leq \frac{U}{2} \\ p(s) & \text{if } v(s) < 50\% \\ U & \text{otherwise} \end{cases}$$

where p(s) is the energy assigned to s by the traditional AFL's original power schedule and U is a maximum energy that can be assigned by AFL. This power schedule implements a hill climbing metaheuristic that always assigns twice the energy to a seed that is at least 50% valid and has an original energy p(s) that is at most half the maximum energy U.

In that case, the aforementioned depth-based power schedule will assign more energy to seeds with a higher degree of validity. It is expected more valid inputs can be generated from already valid inputs. It implements a hill climbing meta-heuristic where the search follows a gradient descent. A seed with a higher degree of validity will always be assigned higher energy than a seed with a lower degree of validity.

Our evaluation demonstrates, against the above 2 innovations, MediaFuzzer can dramatically increase the vulnerability exposure comparing with its baseline AFL. Within the given time limit of 12 hours, against the same test coverage, MediaFuzzer discovered 33 bugs in Intel Media Driver while its baseline (AFL) only detected 5 bugs.

Also, MediaFuzzer is more efficient than AFL.

As shown by [Figure 3] below, against the same integration test, MediaFuzzer can discover the same number of code paths in 9 hours for which AFL requires 24 hours (i.e., about 2.7× more efficient).

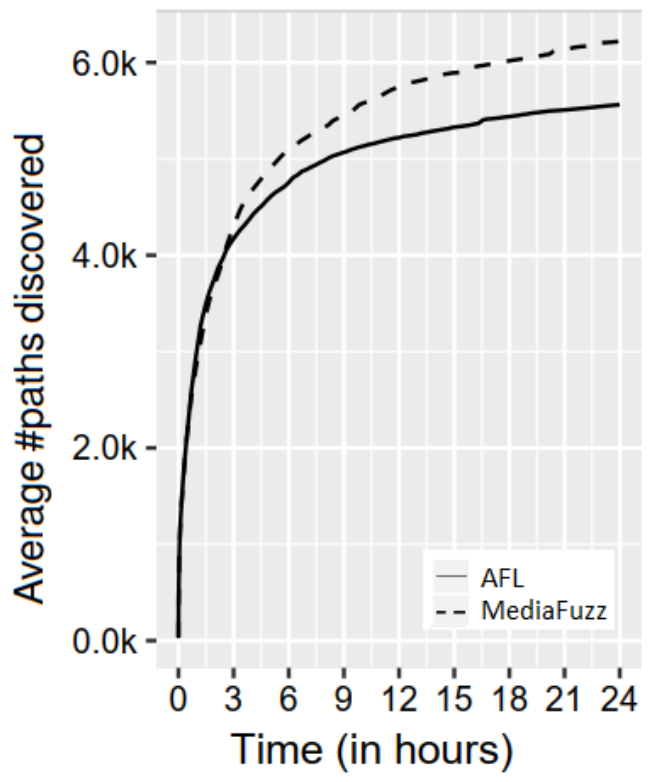

Figure 3. Average #Code Paths discovered in 24 hours for Intel Media driver

Meanwhile its extra overhead is almost negligible. With our optimization, MediaFuzzer now can achieve similar execution speeds to AFL - ±2% for average execution time per seeds.

## 3. Summary

Grey box Fuzzing is a popular and effective approach for vulnerability discovery.

However, conventional Grey box fuzzers like AFL is structure unaware and tend to spend more time on smaller seeds with lower execution time, which greatly impact its fuzzing efficiency for complicated input types.

In this work, we introduce the intelligent grey box input parameter fuzzing, MediaFuzzer, for Intel Media driver. It supports effective fuzzing based on selective fields of input and creates one novel calling chain depth-based power schedule biased toward the seeds corpus leading to deeper calling chain and more likely to pass the parameter validity check.

Per Our evaluation, with the above innovations, MediaFuzzer can dramatically improve the vulnerability exposures (~6.6 times more issues exposed) and fuzzing efficiency (~2.7 times more efficient) than its baseline AFL for Intel media driver with negligible extra overhead.

We believe the similar methodology can also be applied on and benefit other Intel Software.

## ACKNOWLEDGMENTS

We would like to express special thanks of gratitude to our Boss, Xiong Andy, who granted us the great opportunity for this interesting research and get MediaFuzzer applied on our media driver development.

Also, thanks to our Intern, Sun, Weiqi for your help on all kinds of data collection in our research.

**Authors**

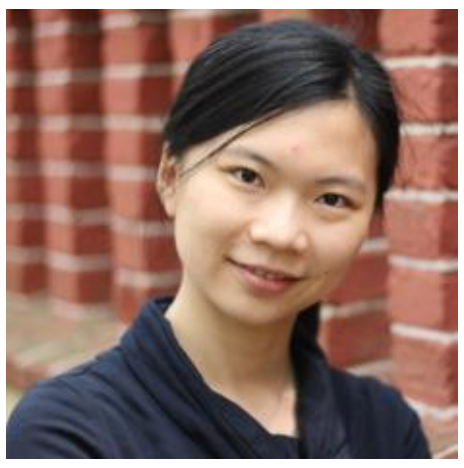

Linlin Zhang is a senior software engineer at Intel. Her research interests include software architecture, quality, continuous delivery, and DevOps. Please contact her at livia.zhang@intel.com

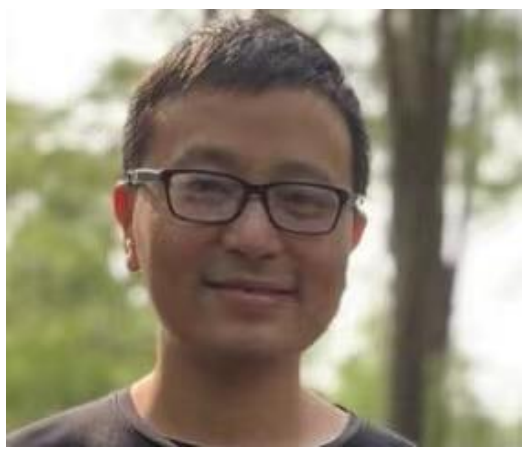

Ning Luo is the senior software architect at Intel. Please contact him at ning.luo@intel.com.